\documentclass[twocolumn,preprintnumbers,amsmath,amssymb]{revtex4-1} 
\usepackage{graphicx}
\usepackage{dcolumn}
\usepackage{amsmath,amsbsy,amssymb,scalefnt}
\usepackage{bm}
\usepackage{epstopdf}
\usepackage{amsfonts}
\usepackage{textcomp}
\usepackage{mathtools}
\usepackage{hyperref}

\hypersetup{backref=true,
pdfnewwindow=true, colorlinks=true,
linkcolor=blue, anchorcolor=blue,
citecolor=blue, filecolor=blue,
menucolor=blue, urlcolor=blue}

\begin{document}
\title{\large\bf $\gamma$-Metrics in Higher Dimensions }

\author{Arash Hajibarat} \thanks{\hyperref{mailto:a.hajibarat@ph.iut.ac.ir}{}{}{a.hajibarat@ph.iut.ac.ir}} \author{Behrouz Mirza} \thanks{\hyperref{mailto:b.mirza@iut.ac.ir}{}{}{b.mirza@iut.ac.ir}} \author{Alireza Azizallahi} \thanks{\hyperref{mailto:a.azizallahi@ph.iut.ac.ir}{}{}{a.azizallahi@ph.iut.ac.ir}}

\address{Department of Physics, Isfahan University of Technology, Isfahan 84156-83111, Iran}

\date{\today}

\begin{abstract} 
We introduce five and higher dimensional $\gamma$-metrics. The higher dimensional metrics are exact solutions of the vacuum field equations and represent new types of singularities. For dimensions $d>5$ we have obtained $\gamma$-metrics in flat coordinates. We obtain  singularities of metrics and for a better understanding of geometrical and physical properties of the five dimensional metric, stable circular orbits are determined by means of the effective potential. Effect of the deformed parameter ($\gamma$) on redshift of the $\gamma$-metrics are calculated. Interior solution for the five-dimensional $\gamma$-metric is also obtained.

\vspace{5mm}
\vspace{5mm}
\end{abstract}
\maketitle

\section{\label{introduction}introduction}

Based on no-heir theorem a black hole can be completely described by it's mass, charge and angular momentum \cite{Misner,Israel1, Israel2, Carter}, and black holes do not have higher multipoles other than monopole. However, exact solutions which have higher non-relativistic multipoles have been found. These solutions with higher multipoles rather than an event horizon have a singular surface and are not geodesically complete and also contradict cosmic censorship hypothesis. We can interpret these new solutions in different ways. One may consider them as solutions which describe gravitational field around an astrophysical compact object without spherical symmetry before formation of a black hole. In this interpretation higher multipoles must be radiated during gravitational collapse. This interpretation is reasonable from this point of view that perfect spherical symmetry is only an abstraction which is not completely available in nature. In another approach toward these solutions, we can dispense with aforementioned difficulties by considering singular surface as a region whereon classical GR is not sufficient for understanding of physics and a coherent theory of quantum gravity is necessary for through understanding of physics in vicinity of singular surface. These objects, for a distant observer, are similar to a black hole due to this fact that singular surface is infinitely redshifted, in other words these solutions mimic a black hole for a distant observer.

One of the interesting solutions with higher non-relativistic multipoles is $\gamma$-metric which was studied by Zipoy and Voorhees, respectively \cite{Zipoy,Voorhees}. In this solution, in addition to mass, there is an extra parameter, shown by $\gamma$, which indicates deviation from spherical symmetry. Singularities of this class of metrics are naked and there is no event horizon. According to the cosmic censorship hypothesis \cite{Penrose}, the existence of a naked singularity without an event horizon is ruled out. However,  naked singularities appear in this type of exact solutions of Einstein's equations and we expect  that both of weak and strong versions of cosmic censorship hypothesis to be  violated by $\gamma$-metric. 

Some aspects of metrics with quadrupoles were investigated in \cite{Francisco}.  Geometrical features of $\gamma$-metric can be found in \cite{Herrera1,Herrera2,Bonnor,Herrera3}. Gravitational lensing of metrics with naked singularity like $\gamma$-metric is investigated in \cite{Virbhadra}.  Interior solutions of four dimensional $\gamma$-metric and axially symmetric metrics, were discussed in \cite{Boshkayev,Hernandez,Stewart}. Motion of massive and massless particles in four dimensional $\gamma$-metric geometry was studied in \cite{Herrera4,Chowdhury1,Benavides,Abdikamalov}. Harmonic oscillation of neutral particles was studied in \cite{Chowdhury2}. Description of spinning particles in $\gamma$-metric and comparison of that with Kerr metric is available in \cite{Toshmatov}.

In this paper, we have obtained the higher dimensional $\gamma$-metrics and explored  some basic properties of them. It is interesting that generalised coefficients in the higher dimensional $\gamma$-metrics indicate that there is no such kind of solutions in two and three dimensional space times.

 This paper is organized as follows. In Sec.~II we introduce a five-dimensional $\gamma$-metric in Hopf spherical coordinates and higher dimensional ones in flat coordinates. We also determine related curvature singularities. In Sec.~III geodesics equations for the five dimensional $\gamma$-metric  are obtained. We also investigate radial motion of both massive and massless particles and obtain elapsed-time for crossing singular surface at $r=\sqrt{2M}$. In Sec.~IV circular orbits for massive and massless particles are determined via calculating the effective potential. We investigate the gravitational redshift and surface gravity in Sec.~V. In Sec.~VI an interior solution for the five dimensional $\gamma$-metric is calculated. Another higher dimensional $\gamma$-metrics in hyperbolic coordinates is introduced in Sec.~VII. Finally, Sec.~VIII is about our concluding remarks. 

All calculations are performed in natural units $G=c=1$.

\section{Higher-Dimensional $\gamma$-Metrics}

Four-dimensional $\gamma$-metric in Erez-Rosen coordinate has been obtained as follows:
\begin{eqnarray}
ds^2=-f^\gamma dt^2+f^{\gamma^2-\gamma}g^{1-\gamma^2}\Big(\frac{dr^2}{f}+r^2d\theta^2\Big)\notag\\+f^{1-\gamma}r^2sin^2\theta~d\phi^2,
\label{A1}
\end{eqnarray}
where 
\begin{eqnarray}
f(r)&=&1-\frac{2M}{r},\notag\\g(r,\theta)&=&1-\frac{2M}{r}+\frac{M^2sin^2(\theta)}{r^2}.
\label{A2}
\end{eqnarray}
This metric is an exact solution of the vacuum field equations and has singularities at $r=0$ and $r=2M$. There is also other singularities at $r=M(1\pm cos(\theta))$ if $\gamma<\sqrt{3/2}$. It should be noted that singularities of this metric are not covered by an event horizon. 

Also by introducing a new coordinate $\rho$ defined by $r=\rho\big(1+\frac{M}{2\rho}\big)^2$, we arrive to the four-dimensional $\gamma$-metric in isotropic form
\begin{eqnarray}
ds^2=-f^\gamma dt^2+\Big(1+\frac{M}{2\rho}\Big)^4\Big[f^\mu k^\nu \big(d\rho ^2+\rho ^2d\theta ^2\big)\nonumber\\+\rho ^2f^\beta sin^2\theta~d\phi ^2\Big],
\label{A3}
\end{eqnarray}
where
\begin{eqnarray}
&& \mu =\gamma(\gamma -1),~~~~\nu =(1-\gamma ^2),~~~~\beta =1-\gamma ,\nonumber\\ &&f=\Big(\frac{M-2\rho}{M+2\rho}\Big)^2,\nonumber\\ &&k=\frac{M^4+16\rho ^4-8M^2\rho ^2cos(2\theta)}{(M+2\rho)^4}.
\label{A4}
\end{eqnarray}
It is interesting to explore higher dimensional $\gamma$-metrics. Study of the $\gamma$-metric in higher dimensions may help us to have a better understanding of singularities in general relativity. Based on four dimensional case, we assume that the five dimensional $\gamma$-metric, that is a solution for the vacuum field equations ($R_{\mu\nu}=0$), would be in the following form

\begin{eqnarray}
	ds^2&=&-f^\gamma dt^2+f^\mu k^\nu\Big(\frac{dr^2}{f}+r^2 d\theta^2\Big)+r^2 f^\beta\big( sin^2\theta~d\phi ^2\nonumber\\&+&cos^2\theta~d\psi ^2\big),\nonumber\\
	f(r)&=&1-\frac{2M}{r^2},\notag\\k(r,\theta)&=&1-\frac{2M}{r^2}+\frac{M^2n^2(\theta)}{r^4},
	\label{A100}
\end{eqnarray}

In this form $\mu,\nu,\beta$ and $n(\theta)$ are unknown parameters and function, respectively. For determining these parameters and function we use $M$ as a perturbation parameter and solve all vacuum field equations ($R_{\mu \nu}=0$)  perturbatively.  At first order of perturbation we have:

\begin{eqnarray}
	R_{rr}&=&\frac{(-6+4\beta+6\gamma+8\mu+8\nu)M}{r^4}+O(M^2),\label{A101}\\
	R_{r\theta}&=&\frac{4(-\beta+\mu+\nu)cos(2\theta)M}{r^3}+O(M^2),\label{A102}\\
	R_{\theta\theta}&=&R_{\phi\phi}=\frac{2(1-\gamma-2\beta)cos^2\theta~ M}{r^2}+O(M^2),\label{A103}
\end{eqnarray}
Therefore
\begin{eqnarray}
	R_{\phi\phi}&=&0~~\rightarrow~~\beta=\frac{1-\gamma}{2},\label{A104}\\
	R_{r\theta}&=&0~~\rightarrow~~\beta=\mu+\nu,\label{A105}\\
	R_{rr}&=&0~~\rightarrow~~\mu=\frac{1-\gamma-2\nu}{2},\label{A106}
\end{eqnarray}
Then using the obtained parameters in the metric we calculate Ricci tensor up to the second order of perturbation parameter $ M $. Therefor we have:
\begin{eqnarray}
	R_{r\theta}&=&\frac{2\nu~ n(\theta)(-2cot(2\theta)n(\theta)+\frac{dn(\theta)}{d\theta})M^2}{r^5}+O(M^3),\label{A107}~~~~~\\
	R_{r\theta}&=&0~~\rightarrow ~~n(\theta)=c_1 sin(2\theta),\label{A108}
\end{eqnarray}
Then we use $ n(\theta)=sin(2\theta) $ in the metric and calculate $ R_{rr} $ as follows:

\begin{equation}
	R_{rr}=\frac{6-6\gamma^2-8\nu}{r^6}M^2+O(M^3),\label{A109}
\end{equation}
Therefore
\begin{equation}
	R_{rr}=0~~\rightarrow~~\nu=\frac{3}{4}(1-\gamma^2).\label{A110}
\end{equation}
Finally we reach the following form for the function and parameters:
\begin{eqnarray}
	n(\theta)&=&sin(2\theta),\label{A111}\\
	\mu&=&\frac{1}{4}(3\gamma+1)(\gamma-1),\label{A112}\\
	\nu&=&\frac{3}{4}(1-\gamma^2),\label{A113}\\
	\beta&=&\frac{1-\gamma}{2}\label{A114}.
\end{eqnarray}
Now we can verify that the following metric is a solution for the vacuum field equations ($R_{\mu\nu}=0$)

\begin{eqnarray}
&&ds^2=-f^\gamma dt^2+f^\mu k^\nu\Big(\frac{dr^2}{f}+r^2 d\theta^2\Big)+r^2 f^\beta\big( sin^2\theta~d\phi ^2\nonumber\\&&+cos^2\theta~d\psi ^2\big),
\label{A5}
\end{eqnarray}
where 
\begin{eqnarray}
f(r)&=&1-\frac{2M}{r^2},\notag\\k(r,\theta)&=&1-\frac{2M}{r^2}+\frac{M^2sin^2(2\theta)}{r^4},\nonumber\\
\mu =\frac{1}{4}(3\gamma +1)(\gamma &-&1),~ \nu = \frac{3}{4}(1-\gamma^2), ~\beta =\frac{1-\gamma}{2}.~~~
\label{A6}
\end{eqnarray}
Through calculation of $R_{\mu\nu\rho\sigma}R^{\mu\nu\rho\sigma}$ for this metric, four singularities at $r=0$, $r=\sqrt{2M}$ and $r=\sqrt{M(1\pm cos(2\theta))}$ are obtained.
The five-dimensional $\gamma$-metric in Eq. ({\ref{A5}}) can be written in the prolate-spheroidal coordinates $(t,x,y,\phi,\psi)$ as follow
\begin{eqnarray}
ds^2 &=& -e^{2\psi_1}dt^2\nonumber\\
&+&\frac{M}{4}e^{2(\psi_3 -\psi_2)}(x^2-y^2)(x+1)^{\delta_4}\Big[\frac{dx^2}{x^2-1}+\frac{dy^2}{1-y^2}\Big]\nonumber\\
&+&\frac{M}{2}e^{-2\psi_2}(x-1)\Big[(1-y)d\phi^2+(1+y)d\psi^2\Big],
\label{A7}
\end{eqnarray}

where
\begin{eqnarray}
&& x=\frac{r^2}{M}-1,~~~~y=cos(2\theta) ,\nonumber\\
&&\psi_1 =\frac{\delta_1}{2}ln(\frac{x-1}{x+1}), ~~~~\psi_2 =\frac{\delta_2}{2}ln(\frac{x-1}{x+1}),\nonumber\\
&&\psi_3 =\frac{\delta_3}{2}ln(\frac{x-1}{x^2-y^2}), \nonumber\\
&&\delta_1 =\gamma , ~~\delta_2 =\frac{\gamma +1}{2},~~\delta_3 =\frac{1}{4}(3\gamma^2 +1),\nonumber\\
&&\delta_4 =\frac{3}{4}(\gamma^2 -1).
\label{A8}
\end{eqnarray}
Also by use of $r=\rho\big(1+\frac{M}{2\rho^2}\big)$ we get to the following five-dimensional $\gamma$-metric in isotropic form
\begin{eqnarray}
ds^2=-f^\gamma dt^2+\Big(1+\frac{M}{2\rho ^2}\Big)^2\Big[f^\mu k^\nu \big(d\rho ^2+\rho ^2d\theta ^2\big)\nonumber\\+\rho ^2f^\beta \big( sin^2\theta~d\phi ^2+cos^2\theta~d\psi ^2\big)\Big],
\label{A9}
\end{eqnarray}
where
\begin{eqnarray}
 f&=&\Big(\frac{M-2\rho ^2}{M+2\rho ^2}\Big)^2,\nonumber\\ k&=&\frac{M^4+16\rho ^8-8M^2\rho ^4cos(4\theta)}{(M+2\rho ^2)^4}.
\label{A10}
\end{eqnarray}

In addition, we introduce two types of d-dimensional flat $\gamma$-metrics and their Kretschmann  scalar as follows:

\subsection{Type 1: $\gamma$-metrics in flat coordinates ($d\geq4$)}

In the following, we obtain a novel class of $\gamma$-metrics in arbitrary higher dimensions. $\gamma$-metrics in flat coordinates have interesting and new type of singularities. We obtained  the following metrics by considering $\mu$, $\nu$ and $\beta$ as arbitrary second order polynomials of parameter $\gamma$ and then solved the vacuum field equations ($R_{\mu \nu}=0$)

\begin{eqnarray}
ds^2&=&-f^\gamma dt^2+f^\mu k^\nu \Big(\frac{1}{f}dr^2+r^2d\theta ^2\Big)+\theta ^2 r^2f^\beta d\phi^2 \nonumber\\&+&\sum_{i=1}^{d-4}r^2f^\beta d\psi _i^2,
\label{A87}
\end{eqnarray}
where $d$ is space-time dimensions and
\begin{eqnarray}
f(r)&=&\frac{2M}{r^{d-3}},~~~~k(r)=\frac{2M}{r^{d-3}}+\frac{(d-3)^2M^2\theta ^2}{r^{2(d-3)}},\nonumber\\ \mu &=&\frac{\gamma -1}{2(d-3)}((d-2)\gamma +d-4),\nonumber\\ \nu &=&\frac{d-2}{2(d-3)}(1-\gamma ^2),~~ \beta =\frac{1-\gamma}{d-3}.
\label{A88}
\end{eqnarray}
The Kretschmann scalar in d-dimensions is 
\begin{eqnarray}
&R&^{\mu \nu \rho \sigma}R_{\mu \nu \rho \sigma}=\frac{x+y}{r^{2(d-1)}\big((d-3)^2M\theta^2 -2r^{(d-3)}\big)}\nonumber\\ &\times&\Big(-\frac{2M}{r^{(d-3)}}\Big)^{-\frac{((d-2)\gamma+ d-4)(\gamma -1)}{d-3}} \nonumber\\ &\times&\Big(\frac{M((d-3)^2M\theta ^2-2r^{(d-3)})}{r^{2(d-3)}}\Big)^{\big(\frac{d-2}{d-3}\big)(\gamma ^2-1)},
\label{A89}
\end{eqnarray} 
where

\begin{eqnarray}
x&=&\frac{M^3\theta ^2(\gamma +1)^2(d-3)^3(d-2)}{2}\nonumber\\ \times\Big((d&-&2)^2\gamma ^4+2(d-2)\gamma ^3+\Big(1+\frac{3(d-3)(d-2)}{2}\Big)\gamma ^2\nonumber\\ -(d&-&4)(d-3)\gamma +\frac{(d-3)(d-4)}{2}\Big),\nonumber\\
\label{A891}
\end{eqnarray}
and
\begin{eqnarray}
y&=&r^{(d-3)}M^2(\gamma +1)^2(d-3)(d-2)\nonumber\\ \Big(&-&\frac{7d^2-31d+36}{2}\gamma ^2+(d-4)(3d-7)\gamma \nonumber\\ &-&\frac{(d-4)(3d-7)}{2}\Big),
\end{eqnarray}

\noindent with singularity at $r=0$. If we replace $M$ with $-M$ we arrive to the following singularities
\begin{eqnarray}
r&=&0,\nonumber\\ r&=&\Big(\frac{(d-3)^2M\theta ^2}{2}\Big)^{\frac{1}{d-3}}~~ for~~~ \bigtriangleup _-<\gamma <\bigtriangleup _+,  
\label{A90}
\end{eqnarray}
where
\begin{equation}
\bigtriangleup_\pm=\frac{1}{2}\Big(1\pm\sqrt{\frac{5d-4}{d-2}}\Big).
\label{A91}
\end{equation}

\subsection{Type 2: $\gamma$-metrics in toroidal coordinates}

In this part, we introduce another new class of  higher dimensional $\gamma$-metrics which are exact  solutions of the vacuum field equations ($R_{\mu \nu} =0$)  with a novel type of singularities. This is the simplest kind of $\gamma$-metrics where singularity appears only at $r=0$. However the order of singularities are  different from singularities of the corresponding  black holes ($\gamma=1$). 

\begin{eqnarray}
ds^2&=&-f^\gamma dt^2+f^\mu \Big(\frac{1}{f}dr^2+r^2 d\phi ^2\Big)\nonumber\\&+&\sum_{i=1}^{d-3}r^2f^\beta d\psi _i^2,
\label{A92}
\end{eqnarray}
where
\begin{eqnarray}
f(r)&=&\frac{2M}{r^{d-3}},\nonumber\\ \mu &=&\frac{d-2\gamma-(d-2)\gamma ^2}{2(d-3)}, ~~\beta =\frac{1-\gamma}{d-3}.
\label{A93}
\end{eqnarray}

The Kretschmann  scalar in d-dimensions is

\begin{equation}
R^{\mu \nu \rho \sigma}R_{\mu \nu \rho \sigma}=\frac{z}{r^{2(d-1)}}\Big(\frac{2M}{r^{(d-3)}}\Big)^{\frac{((d-2)\gamma +d)(\gamma -1)}{d-3}},
\label{A94}
\end{equation}

where

\begin{eqnarray}
z&=&\frac{M^2(\gamma+1)^2(d-3)(d-2)}{2}\big((d-2)^2\gamma ^4\nonumber\\&+&2(d-2)\gamma ^3+\big(1+\frac{3(d-3)(d-2)}{2}\big)\gamma ^2\nonumber\\ &-&(d-4)(d-3)\gamma +\frac{(d-4)(d-3)}{2}\big),
\label{A941}
\end{eqnarray}

\noindent with singularity at $r=0$.

In the following sections we study some properties of the five dimensional $\gamma$-metric and also introduce other types of $\gamma$-metrics in higher dimensions.

\section{Analyzing $\gamma$-metric geodesics}
In this section, we study geodesics in the five dimensional $\gamma$-metric space-time. We consider behavior of geodesics near $r= \sqrt{2M}$. One of the important parameters in understanding the geometry of a space-time is test particles’ elapsed-time for passing specific points on a manifold. In the following, we will calculate elapsed-time near the curvature singularity by using geodesics equations for a radially falling particle. 

\subsection*{Massive particles}
It is important to study radial motion of a test particle falling toward the singularity at $r=\sqrt{2M}$. For simplicity, we assume the motion is taking place on equatorial plane $\theta=\frac{\pi}{2}$. The four velocity of the test particle is $u^\mu=\frac{dx^\mu}{d\lambda}$ in which $\lambda$ is an affine parameter. In the five dimensional space-time for radial motion of the particle we have $u^2=u^3=u^4=0$. The motion of this particle is governed by the following equation 
\begin{eqnarray}
\frac{du^0}{d\lambda}&=&-\Gamma^0_{\mu\nu}u^\mu u^\nu = -g^{00}\frac{dg_{00}}{d\lambda}u^0.
\label{A14}
\end{eqnarray}

\noindent Eq. ({\ref{A14}}) can be written as below:
\begin{equation}
g_{00}\frac{du^0}{d\lambda}+\frac{dg_{00}}{d\lambda}u^0=\frac{d}{d\lambda}\big(g_{00}u^0\big)=0,
\label{A15}
\end{equation}
Therefore $g_{00}u^0$ is a constant. At this point by considering normalization condition, $g_{\mu\nu}u^\mu u^\nu=-1$, $u^1$ can be obtained as follows: 
\begin{equation}
u^1=-\Big(1-\frac{2M}{r^2}\Big)^\frac{1-\gamma}{4}\bigg(h^2-\Big(1-\frac{2M}{r^2}\Big)^\gamma\bigg)^{\frac{1}{2}},
\label{A17}
\end{equation} 
where $h=g_{00}u^0$. We may write $u^0/u^1$ as below
\begin{equation}
\frac{dt}{dr}=\frac{u^0}{u^1}=h\Big(1-\frac{2M}{r^2}\Big)^\frac{-(3\gamma+1)}{4}\bigg(h^2-\Big(1-\frac{2M}{r^2}\Big)^\gamma\bigg)^{-\frac{1}{2}}.
\label{A18}
\end{equation}
For analyzing motion of test particles near the singularity $r=\sqrt{2M}$, we assume $\epsilon=r-\sqrt{2M}$, where $\epsilon$ is a very small number such that $\epsilon^n$ for $n\geq2$ is negligible and Eq. ({\ref{A18}}) can be written as
\begin{equation}
\frac{dt}{dr}=-\bigg(\frac{\sqrt{2M}}{2(r-\sqrt{2M})}\bigg)^\frac{3\gamma+1}{4}.
\label{A19}
\end{equation}
Now integration of ({\ref{A19}}) for $\gamma=1$ yields
\begin{equation}
 t=-\sqrt{\frac{M}{2}}log\big(r-\sqrt{2M}\big)+constant,
\label{A20}
\end{equation}
and integration of ({\ref{A19}}) for $\gamma\neq1$ is given by 
\begin{equation}
 t=-\Big(\frac{4}{3(\gamma-1)}\Big)\bigg(\frac{M}{2}\bigg)^\frac{3\gamma+1}{8}\Big(r-\sqrt{2M}\Big)^\frac{3(1-\gamma)}{4}+constant.
\label{A21}
\end{equation}
Eq.({\ref{A21}}) shows for $\gamma>1$ when a test particle approaches $r=\sqrt{2M}$, $t$ is divergence and for $\gamma<1$, $t$ converges to a finite value which is in contrast to $\gamma=1$ case in which infinite time was measured by a distant observer. For a comoving observer we have:
\begin{equation}
\frac{d\tau}{dr}=\frac{1}{u^1}=-\Big(1-\frac{2M}{r^2}\Big)^\frac{(\gamma-1)}{4}\bigg(h^2-\Big(1-\frac{2M}{r^2}\Big)^\gamma\bigg)^{-\frac{1}{2}},
\label{A22}
\end{equation}
where $\tau$ is the proper time and near to the singularity ($r= \sqrt{2M}$) can be written as 
\begin{equation}
\tau =-\frac{\sqrt{M}2^{\frac{1}{8}(17-\gamma)}}{h(\gamma+3)}\Big(\frac{r-\sqrt{2M}}{\sqrt{M}}\Big)^\frac{\gamma +3}{4}+constant.
\label{A23}
\end{equation}

This relation shows the elapsed time for comoving observer is always finite if $\gamma>0$. After reiterating the above calculations for four-dimensional $\gamma$-metric ({\ref{A1}}), we obtain 
\begin{eqnarray}
t &=&-2Mlog(r-2M)+constant \label{A24}~~for~~\gamma =1,\\
t &=& \frac{2M}{\gamma -1}\big(r-2M\big)^{1-\gamma}+constant \label{A25}~for~\gamma \neq 1,\\
\frac{d\tau}{dr} &=& h^{-1} \label{A26} ~~~~~~~~~~~~~~~~~~~~~~~~~~~~~~~~~~~for~~\gamma > 0,
\end{eqnarray}
These results indicate that $\gamma>1$ leads to infinite elapsed time for a distant observer,
however elapsed time is finite for $\gamma<1$.

\subsection*{Massless particles}
We study radial null geodesic of massless particles near the singularity at $r=\sqrt{2M}$. We assume $(\phi,\psi)$ are constants and for simplicity, as before restrict motion on equatorial plane $\theta=\frac{\pi}{2}$. As a result of these constraints, we reach
\begin{equation}
ds^2=-\Big(1-\frac{2M}{r^2}\Big)^\gamma dt^2+\Big(1-\frac{2M}{r^2}\Big)^{-\frac{\gamma +1}{2}}dr^2=0,
\label{A27}
\end{equation}
which describe radial null geodesics and may be written as
\begin{equation}
\frac{dt}{dr}=\pm\Big(1-\frac{2M}{r^2}\Big)^{-\frac{3\gamma +1}{4}}.
\label{A28}
\end{equation}
In agreement with our expectation for a flat metric, the slope of light-cones goes to $+1$ when $r$ goes to infinity (for all values of $\gamma$). In addition near the singularity, ({\ref{A28}}) reduces to ({\ref{A19}}) which means that massive and massless particles have similar behaviors.

\section{Orbits of test particles}
In this section for a better understanding of physical features of  $\gamma$-metric, we determine circular orbits for test particles by means of calculating the effective potential. We use conserved quantities of the five dimensional $\gamma$-metric for calculating the circular geodesics.

\subsection{Effective potential}
In this section our goal is calculation of effective potential via two slightly different methods, using Hamiltonian formalism for massive particles and Killing vectors for massless particles of the metric.

\subsection*{Massive particles}
Hamilton formalism can be used to determine circular orbits for massive particles. Hamiltonian of a massive particle with mass $m$ in a curved space-time is as below
\begin{equation}
H=\frac{1}{2}g_{\mu\nu}p^\mu p^\nu +\frac{1}{2}m^2,
 \label{A29}
\end{equation}
where $p^\mu$ is four momentum of the test particle which can be defined as $p^\mu=mu^\mu$. Here $u^\mu$ is four velocity of the test particle.
Because our metric is explicitly independent of $ t,\phi$ and $\psi$, Hamiltonian ({\ref{A29}}) is also independent of these parameters. As a result, their correspondent canonical momenta are conserved quantities. The energy and angular momenta are as below
\begin{eqnarray}
p_t&=&g_{tt}\frac{dt}{d\tau}=-E, ~~~~p_\phi= g_{\phi\phi}\frac{d\phi}{d\tau}=L_\phi,\nonumber\\ p_\psi &=&g_{\psi\psi}\frac{d\psi}{d\tau}=L_\psi,
 \label{A30}
\end{eqnarray}
in which $E,L_\phi$ and $L_\psi$ are constants and respectively known as total energy, angular momentum around $\phi$ axis and angular momentum around $\psi$ axis.
Now we rewrite Hamiltonian in Eq.({\ref{A29}}) as follows   
\begin{equation}
H=\frac{1}{2}g^{rr}p_r^2+\frac{1}{2}g^{\theta\theta}p_\theta ^2+H_{t\phi\psi},
 \label{A31}
\end{equation}
wherein $H_{t\phi\psi}$ is defined in
\begin{equation}
H_{t\phi\psi}(r,\theta)=\frac{1}{2}\big(g^{tt}E^2+g^{\phi\phi}L_\phi^ 2+g^{\psi\psi}L_\psi ^2+m^2\big).
 \label{A32}
\end{equation}
By considering this fact that normalization condition, $g_{\mu\nu}u^\mu u^\nu= -1$, for massive particles with unit mass, leads to $H=0$, Eq. ({\ref{A31}}) will be resulted in 
\begin{equation}
g_{rr}\big(\frac{dr}{d\tau}\big)^2+g_{\theta\theta}\big(\frac{d\theta}{d\tau}\big)^2=-\frac{2H_{t\phi\psi}(r,\theta)}{m^2}.
 \label{A33}
\end{equation}

As before for simplicity, we assume $\theta=\pi/2$. This constraint lead to $g_{\psi\psi}=0$ and then because of ({\ref{A30}}) we have 
\begin{equation}
g_{rr}\big(\frac{dr}{d\tau}\big)^2=-\frac{2H_{t\phi\psi}(r,\theta)}{m^2}.
 \label{A34}
\end{equation}

\noindent Using Eq. ({\ref{A32}}), we rewrite ({\ref{A34}}) as follows
\begin{equation}
\frac{1}{2}f^{-\beta}\big(\frac{dr}{d\tau}\big)^2+V_{eff}(r)=\xi,
 \label{A35}
\end{equation}
where
\begin{eqnarray}
 \xi  &\equiv&\frac{E^2}{2},\label{A37}\\
V_{eff}(r)&=&\frac{1}{2}\Big(1-\frac{2M}{r^2}\Big)^\gamma +\frac{L_\phi ^2}{2r^2}\Big(1-\frac{2M}{r^2}\Big)^{\frac{3\gamma -1}{2}}, \label{A36}
\end{eqnarray}
which is mathematically analogous to the energy of a particle with unit mass and potential $V_{eff}(r)$ in one dimension.

\subsection*{Massless particles}
In this case, we use symmetries of the five dimensional $\gamma$-metric ({\ref{A5}}) for deriving effective potential. Due to this fact that the metric is explicitly independent of $t,\phi$ and $\psi$, our $\gamma$-metric has three symmetries and three resulted Killing vectors which are demonstrated below
\begin{eqnarray}
K^\mu =\big(1,0,0,0,0\big), \label{A38}\\
R^\mu =\big(0,0,0,1,0\big), \label{A39}\\
S^\mu =\big(0,0,0,0,1\big). \label{A40}
\end{eqnarray}
Any Killing vector $K_\mu$ lead to a conserved quantity $K_\mu \frac{dx^\mu}{d\lambda}$. Here we have three conserved quantity related to Killing vectors. In our notation $K^\mu$ is a time like Killing vector that corresponds to conservation of energy.  $R^\mu$ and $S^\mu$ are space like Killing vectors related to angular momentum conservation around $\phi$ and $\psi$ axes, respectively. By lowering indices of Killing vectors, we can determine conserved quantities explicitly as 
\begin{eqnarray}
E&=&-K_\mu\frac{dx^\mu}{d\lambda}=\Big(1-\frac{2M}{r^2}\Big)^\gamma \big(\frac{dt}{d\lambda}\big), \label{A41}\\
L_\phi &=&R_\mu\frac{dx^\mu}{d\lambda}=r^2\Big(1-\frac{2M}{r^2}\Big)^{\frac{1-\gamma}{2}} \big(\frac{d\phi}{d\lambda}\big), \label{A42}\\
L_\psi &=&S_\mu\frac{dx^\mu}{d\lambda}=0,  \label{A43}
\end{eqnarray}
It is well known from differential geometry that quantity $\varepsilon=-g_{\mu\nu}\frac{dx^\mu}{d\lambda}\frac{dx^\nu}{d\lambda}$ remains constant along a geodesic line, therefore by expanding this quantity we have
\begin{eqnarray}
&&-\Big(1-\frac{2M}{r^2}\Big)^\gamma\big(\frac{dt}{d\lambda}\big)^2+\Big(1-\frac{2M}{r^2}\Big)^{-\frac{\gamma+1}{2}}\big(\frac{dr}{d\lambda}\big)^2\nonumber\\
&&+r^2\Big(1-\frac{2M}{r^2}\Big)^{\frac{1-\gamma}{2}}\big(\frac{d\phi}{d\lambda}\big)^2=-\varepsilon,  \label{A44}
\end{eqnarray}
in which $\varepsilon$ is a constant. By applying equations ({\ref{A41}}), ({\ref{A42}}) and ({\ref{A43}}) to ({\ref{A44}}) we will have
\begin{eqnarray}
&&-E^2+\Big(1-\frac{2M}{r^2}\Big)^{\frac{\gamma -1}{2}}\big(\frac{dr}{d\lambda}\big)^2+\frac{L_\phi ^2}{r^2}\Big(1-\frac{2M}{r^2}\Big)^{\frac{3\gamma-1}{2}}\nonumber\\
&&+\varepsilon\Big(1-\frac{2M}{r^2}\Big)^\gamma =0.
 \label{A45}
\end{eqnarray}

Eq. \noindent ({\ref{A45}}) can be abbreviated as
\begin{equation}
\frac{1}{2}f^{-\beta}\big(\frac{dr}{d\lambda}\big)^2+V_{eff}(r)=\xi,
 \label{A46}
\end{equation}
where $\xi =\frac{E^2}{2}$ and
\begin{eqnarray}
&V&_{eff}(r)=\frac{\varepsilon}{2}\Big(1-\frac{2M}{r^2}\Big)^\gamma +\frac{L_\phi ^2}{2r^2}\Big(1-\frac{2M}{r^2}\Big)^{\frac{3\gamma -1}{2}}, \label{A47}
\end{eqnarray}
For massless particles, which move on null curves, we have $\varepsilon=0$. On the other hand, for massive particles by imposing the normalization condition, which is equivalent to $\varepsilon=1$, we obtain the earlier result in Eq. ({\ref{A36}}). 

\subsection{Circular orbits}

In this section we investigate circular orbits for the $\gamma$-metric, but prior to that we evaluate specific case of $\gamma= 1$ which is five-dimensional Schwarzschild metric. In this specific case ({\ref{A47}}) yields
\begin{equation}
V_{eff}(r)=\frac{\varepsilon}{2}-\frac{M\varepsilon}{r^2}+\frac{L_\phi ^2}{2r^2}-\frac{ML^2}{r^4},
 \label{A49}
\end{equation}
in which the first term is a constant, the second term is five-dimensional Newtonian potential, the third term is centrifugal potential and last one is a pure GR effect. The equation ({\ref{A47}}) has a circular solution at $r=r_0$ if and only if the following equations
\begin{eqnarray}
V_{eff}(r_0)=\xi, \label{A50}\\
\partial _rV_{eff}(r_0)=0,  \label{A51}
\end{eqnarray}
are satisfied. By solving the above equations for massive case ($\varepsilon=1$), we have
\begin{eqnarray}
&&L_\phi =\pm r\Big(1-\frac{2M}{r^2}\Big)^{\frac{1-\gamma}{4}}\sqrt{\frac{2\gamma M}{r^2-3\gamma M-M}}, 
\label{A52}\\
&&E=\Big(1-\frac{2M}{r^2}\Big)^{\frac{\gamma}{2}}\sqrt{\frac{r^2-\gamma M-M}{r^2-3\gamma M-M}}.
 \label{A53}
\end{eqnarray}
 Demanding positivity of terms under radicals yields the following conditions
\begin{eqnarray}
r&>&\sqrt{M(3\gamma +1)} ~~~~  if~~~~~~~~~\gamma\geq\frac{1}{3},
 \label{A54}\\
r&>&\sqrt{2M} ~~~~~~~~~~~~~if  ~~~~ 0<\gamma <\frac{1}{3},
 \label{A55}
\end{eqnarray}
For having stable circular orbits, our solutions must satisfy another stability condition which is $V''_{eff}(r)>0$. We define the innermost stable circular orbit (ISCO) as solutions for $V''_{eff}(r)=0$, which is given by
\begin{equation}
r_{isco}=\sqrt{\frac{(\gamma+1)(3\gamma+1)M}{4\gamma}},
 \label{A56}
\end{equation}
 The stability condition ($V_{eff}''(r)>0$) with ({\ref{A54}}) and ({\ref{A55}}), yields
\begin{eqnarray}
&&r\in(\sqrt{M\big(3\gamma +1)},\infty\big) ~~~~for~~~~ \gamma\in\big[\frac{1}{3},\infty\big),
 \label{A57}\\
&&r\in\big(\sqrt{2M},r_{isco}\big)~~~~~~~~~~~  for~~~~  \gamma\in\big(0,\frac{1}{3}\big),
 \label{A58}
\end{eqnarray}
which determine where stable orbits do exist.
For massless particles ($\varepsilon=0$), effective potential ({\ref{A47}}) reduces to
\begin{equation}
V_{eff}(r)=\frac{L_\phi ^2}{2r^2}\Big(1-\frac{2M}{r^2}\Big)^{\frac{3\gamma -1}{2}}.
 \label{A59}
\end{equation}
Therefore equations in ({\ref{A50}}) and ({\ref{A51}}) only can determine the ratio of $E$ and $L$. The effective potential in (\ref{A59}) yields only one unstable circular orbit for massless particles as follows
\begin{equation}
r=\sqrt{M(3\gamma+1)}~~~~ with ~~~~ \gamma >\frac{1}{3}~~ and ~~\gamma \neq 2n+1,
 \label{A60}
\end{equation}
 where $n\in \mathbb{N}$. Figure (1) shows this massless potential. 
 \begin{figure}
 \centering 
{\includegraphics[angle=0,width=0.3\textwidth]{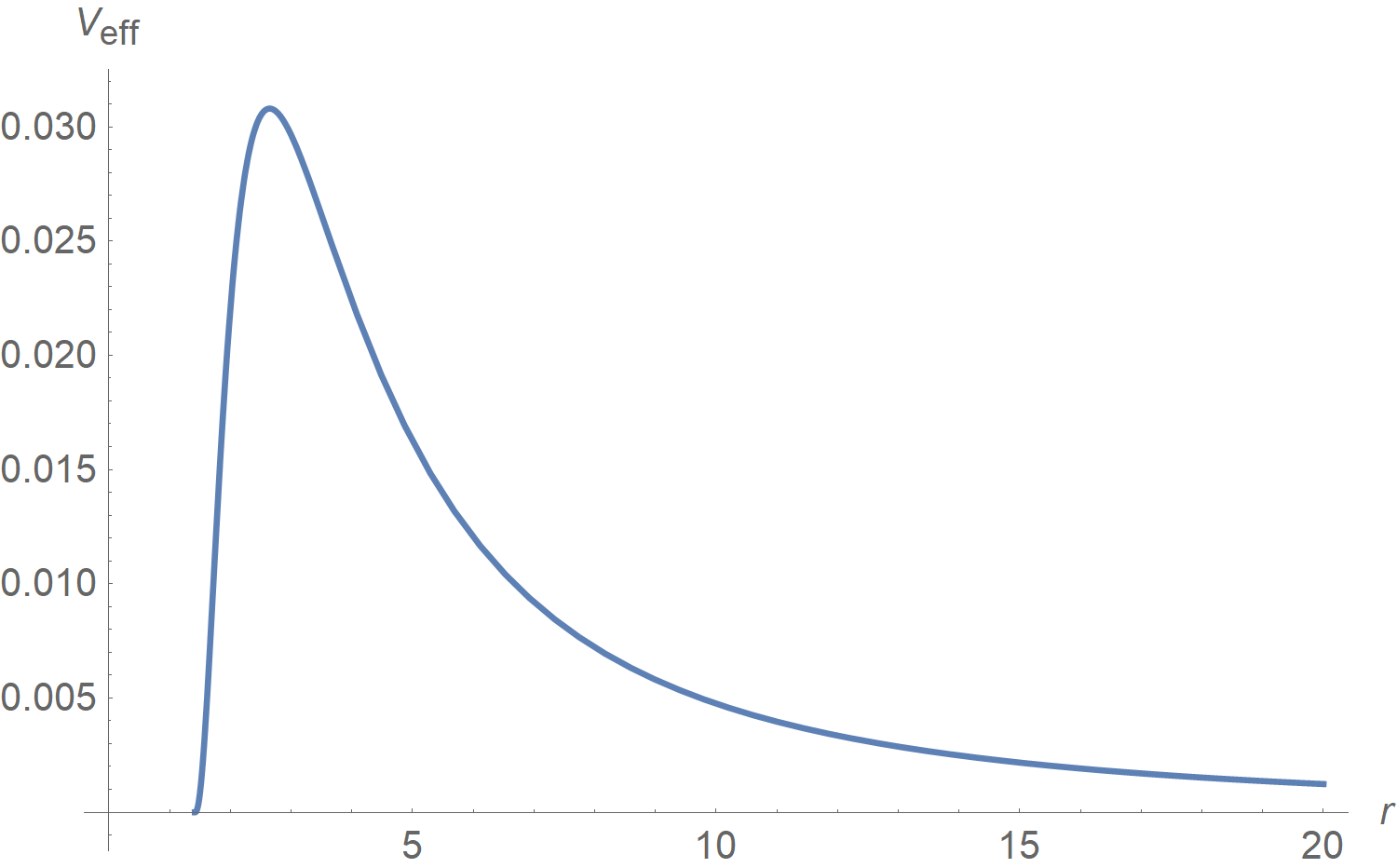}}
\caption{Effective potential ({\ref{A59}}) for massless particles} 
 \end{figure}
 \section{Comparison of $\gamma$-metrics and black holes}
 In this part we compare some aspects of $\gamma$-metrics with black holes. First we calculate gravitational redshift for $\gamma$-metric. For this goal we consider a static observer with four velocity $u^\mu$, and $u^i=0$. Using the normalization condition $g_{\mu\nu}u^\mu u^\nu=-1$, we reach to $u^0=\big(1-2M/r^2\big)^{-\frac{\gamma}{2}}$. This observer measures the frequency of an emitted photon which move on a null geodesic as 
 \begin{equation}
 \Omega =-g_{\mu\nu}\frac{dx^\mu}{d\tau}\frac{dx^\nu}{d\lambda}=-g_{\mu\nu}u^\mu\frac{dx^\nu}{d\lambda}.
  \label{A61}
 \end{equation}
 After imposing ({\ref{A41}}) on ({\ref{A61}}) we obtain
 \begin{equation}
\Omega=- g_{00}u^0\frac{dx^0}{d\lambda}=\Big(1-\frac{2M}{r^2}\Big)^{\frac{\gamma}{2}}\frac{dt}{d\lambda}=\Big(1-\frac{2M}{r^2}\Big)^{-\frac{\gamma}{2}}E.
 \label{A62}
 \end{equation}
We assume that an emitting atom to be  at $r_1$ and the observer at $r_2$. The measured frequency at $r_2$ is as below
 \begin{equation}
 \Omega _2=\bigg(\frac{\Lambda _1}{\Lambda _2}\bigg)\Omega _1.
  \label{A63}
 \end{equation}
 where $\Lambda_i=\big(1-2M/r^2_i\big)^{\frac{\gamma}{2}}$ is redshift factor and $\Omega_1$ and $\Omega_2$ are frequencies at $r_1$ and $r_2$ respectively. In large radii approximation $r_1,r_2>>\sqrt{2M}$, we reach 
 \begin{equation}
 \Omega _2=\Big(1+\gamma \Big(-\frac{M}{r_1^2}+\frac{M}{r_2^2}\Big)\Big)\Omega _1,
  \label{A64}
 \end{equation}
 which shows  for $\gamma>1$ and $\gamma<1$ the redshift, compared to Schwarzschild case, will be increased and decreased respectively. For calculating surface gravity and then temperature, at first by considering $U^\mu=\Big(\big(1-2M/r^2\big)^{-\frac{\gamma}{2}},0,0,0,0\Big)$, $K^\mu=\Lambda U^\mu$  and Killing vector ({\ref{A38}}), we read redshift factor $\Lambda=\big(1-2M/r^2\big)^{\frac{\gamma}{2}}$. Thereafter by means of $a_\mu=\nabla _\mu Ln\Lambda$ acceleration can be calculated as ($a=\sqrt{a_\mu  a^\mu}$)
 \begin{equation}
a=\frac{2M\gamma}{r^3\Big(1-\frac{2M}{r^2}\Big)^{\frac{\mu+1}{2}}\Big(1-\frac{2M}{r^2}+\frac{M^2sin^2(2\theta)}{r^4}\Big)^{\frac{\nu}{2}}}.
 \label{A65}
 \end{equation}
 At last putting redshift factor and acceleration in $\kappa=\Lambda a$ and use $T=\frac{\kappa}{2\pi}$, we obtain the following form
 \begin{equation}
T =\frac{M\gamma}{\pi r^3\Big(1-\frac{2M}{r^2}\Big)^{\frac{3}{8}(\gamma-1)^2}\Big(1-\frac{2M}{r^2}+\frac{M^2sin^2(2\theta)}{r^4}\Big)^{\frac{3}{8}(1-\gamma^2)}}.
 \label{A66}
 \end{equation}
It should be noted that $T$ at $r=\sqrt{2M}$ is infinite for $\gamma\neq1$ which means that we can not define a well defined temperature for the $\gamma$-metric with naked singularity. However at $\gamma=1$ we have a well defined temperature that corresponds to the Schwarzschild black hole. This illustrates that thermodynamical quantities such as temperature and entropy are not well-defined for the $\gamma$-metrics at least based on semiclassical approximations and a complete and consistent quantum theory of gravity is necessary for attributing appropriate and well-defined thermodynamical quantities to the $\gamma$-metrics with naked singularities. Therefore undefinability of temperature and entropy for this metric is another substantial difference between black holes and $\gamma$-metrics.
 
 \section{an interior solution}
 
Interior solution for four dimensional $\gamma$-metric was investigated in \cite{STEWART2}. In this section we obtain an interior solution for our five dimensional $\gamma$-metric. Interior solution just means a star with no singularity that is defined by well defined  energy density and pressure and its radius is greater than $\sqrt{2M}$. The gravitational field outside of the star is given by the $\gamma$-metric. Instead  of directly solving Einstein equations with a source, we ansatz  an interior solution which satisfies boundary conditions and in special case of $\gamma=1$ reduces to interior solution of five dimensional Schwarzschild case \cite{Krori,Shen,Ponce}. For having a well-behaved solution on the boundary we require the continuity of first and second fundamental forms. In addition to these boundary conditions the Synge's junction condition also must be satisfied \cite{Synge}. It can be shown that all boundary and junction conditions are satisfied if  $g_{00}$, $g_{11}$, $g_{22}$, $g_{33}$, $g_{44}$ and their first derivatives are continuous. We consider five dimensional interior solution and observed that in this case interior solution can be obtained from the exterior one by the following transformation 

\begin{equation}
\frac{M}{r^2}\longrightarrow \frac{Mr^2}{R^4},
\label{A67}
\end{equation}

\noindent for components that continuity of the first derivative is not necessary. For other components that the first derivatives are continuous, we use the following transformation that give us interior solution from the exterior one:

\begin{equation}
\frac{2M}{r^2}\longrightarrow 1-\frac{1}{4}\Big(4 \sqrt{1-\frac{2M}{R^2}}-2\sqrt{1-\frac{2Mr^2}{R^4}}\Big)^2,
\label{A68}
\end{equation}

We have imposed those transformations on the five-dimensional $\gamma$-metric and for $g_{rr}$ a mixture of ({\ref{A67}}) and ({\ref{A68}}) for having a healthy behavior at $r=0$. Then we have verified aforementioned continuity conditions on the metric on the boundary at $r=R$. The final metric can be written as follows 

\begin{eqnarray}
ds^2&=&-f^{ \gamma }dt^2+\Big(\frac{\Sigma}{\Delta}\Big)^\nu f^{\beta}\Big(\frac{dr^2}{\Delta}+r^2 d\theta^2 \Big)\nonumber\\ &+& r^2 f^{\beta }\big( \sin ^2\theta d\phi ^2 + \cos ^2\theta d\psi ^2 \big),
\label{A69}
\end{eqnarray}

\noindent where

\begin{eqnarray}
f&=&\frac{1}{4}\Big(4 \sqrt{1-\frac{2M}{R^2}}-2\sqrt{1-\frac{2Mr^2}{R^4}}\Big)^2,\label{A70} \\ \Delta &=&1-\frac{2Mr^2}{R^4},\label{A71}\\ \Sigma &=&1-\frac{2Mr^2}{R^4}+\frac{M^2r^4 \sin ^2(2 \theta )}{R^8},\label{A72}\\
\nu&=&\frac{3}{4} (1-\gamma^2),~~\beta=\frac{1-\gamma}{2}.\label{A73}
\end{eqnarray}

 The interior solution of $\gamma$-metric in ({\ref{A69}}) satisfies all boundary and junction conditions and at $\gamma=1$ reduces to the interior solution of five dimensional Schwarzschild black hole. By using the Einstein equations we can calculate components of  energy momentum tensor as follows: 
 
 \begin{eqnarray}
8\pi T^0_{~0}&=&\frac{12}{R^4}\Big[\gamma M+\Big(\big(r^2(\gamma-2)\nonumber\\ &-&2R^2(\gamma-1)\big)(\gamma-1)\Big)\frac{M^2}{R^4}\Big],\label{A75}\\
-8\pi T^1_{~1}&=&\frac{3}{R^8}\Big[4R^2-r^2(\gamma^2+3)\nonumber\\
&+&r^2(\gamma^2-1)cos(4\theta)\Big]M^2,\label{A76}\\
8\pi T^1_{~2}&=&-\frac{3}{R^8}\Big[r(\gamma^2-1)sin(4\theta)\Big]M^2,\label{A77}\\
-8\pi T^2_{~2}&=&\frac{3}{R^8}\Big[4R^2+r^2(\gamma^2-5)\nonumber\\&-&r^2(\gamma^2-1)cos(4\theta)\Big]M^2,\label{A78}\\
-8\pi T^3_{~3}&=& \frac{12}{R^8}\Big[R^2-r^2 \Big]M^2,\label{A79}\\
-8\pi T^4_{~4}&=&\frac{12}{R^8}\Big[R^2-r^2 \Big]M^2,~~\label{A80}
\end{eqnarray}
 
\noindent where  for simplicity we have expanded results around  $M$ and keep terms up to order two. For sufficiently small $M$ the achieved energy momentum tensor is satisfying $T^0_0>0, T^1_1<0 , T^2_2<0, T^3_3 <0$  which are essential for interior solution. The interior solution does not have singularity and represent a star with multipole moments. For $ 0<\gamma<1 $ and $ \gamma>1 $ we have a prolate and oblate configurations respectively. It should be noted that $T^1_2$ is not zero which means that the interior solution is a star with multipole moments. The mass distribution is not exactly spherically symmetric and depends on both $r$ and $\theta$.

 \section{hyperbolic $\gamma$-metrics}
 
 We also have obtained hyperbolic $\gamma$-metric in four and five dimensions. These four and five dimensional $\gamma$-metrics are exact solutions of the vacuum field equations ($ R_{\mu \nu}=0 $) and represent new types of singularities which is related to hyperbolic functions.
 
\subsection{Four dimensions}
 
 Four dimensional hyperbolic  $\gamma$-metric can be written as follows

 \begin{eqnarray}
ds^2=-f^\gamma dt^2+f^\mu k^\nu\Big(\frac{dr^2}{f}&+&r^2 d\theta^2\Big)\nonumber\\&+&r^2 f^\beta sinh^2\theta~d\phi^2,
\label{A81}
\end{eqnarray}

\noindent where 

\begin{eqnarray}
f(r)&=&-1+\frac{2M}{r},\notag\\k(r,\theta)&=&-1+\frac{2M}{r}+\frac{M^2sinh^2\theta}{r^2},\nonumber\\
\mu &=&\gamma ^2-\gamma ,~~ \nu = 1-\gamma ^2, ~~\beta =1-\gamma ,
\label{A82}
\end{eqnarray}

\noindent with singularities:
\begin{eqnarray}
r&=&0~~~~~~~~for~~~\gamma >0,\nonumber\\
r&=&2M~~~~for~~~~\gamma>1.
\label{A83}
\end{eqnarray}

\subsection{Five dimensions}

We have obtained the five dimensional hyperbolic $\gamma$-metric which is an exact solution of Einsteins equations as follows:
 
\begin{eqnarray}
&&ds^2=-f^\gamma dt^2+f^\mu k^\nu\Big(\frac{dr^2}{f}+r^2 d\theta^2\Big)+r^2 f^\beta\big( sinh^2\theta~d\phi^2\nonumber\\&&+cosh^2\theta~d\psi^2\big),
\label{A84}
\end{eqnarray}

\noindent where 
\begin{eqnarray}
f(r)&=&-1+\frac{2M}{r^2},\notag\\k(r,\theta)&=&-1+\frac{2M}{r^2}+\frac{M^2sinh^2(2\theta)}{r^4},\nonumber\\
\mu =\frac{1}{4}(3\gamma +1)(\gamma &-&1),~ \nu = \frac{3}{4}(1-\gamma^2), ~\beta =\frac{1-\gamma}{2}.~~~~
\label{A85}
\end{eqnarray}

\noindent Using the Kretschmann scalar we have found the following singularities for the five dimensional $\gamma$-metric:
\begin{eqnarray}
r&=&0~~~~~~~~~~~~~~~~~~~~~~~~~~~~~~for~~~~~\gamma >0,\nonumber\\
r&=&\sqrt{M(1\pm cosh(2\theta))}~~~~~~~for~~~~\gamma <1,\nonumber\\
r&=&2M~~~~~~~~~~~~~~~~~~~~~~~~~~~for~~~~\gamma>0.
\label{A86}
\end{eqnarray}

\section{CONCLUSION}

In this paper we introduced higher dimensional $\gamma$-metrics. Our main goal was to obtain exact  solutions of the vacuum Einstein's field equations with new types of singularities. It is expected that classical singularities in general relativity should be resolved in a quantum theory of gravity. We believe that  semi-classical methods might be useful for finding a solution for classical singularities in general relativity. We calculated elapsed-time that observed by a distant observer for a falling particle to cross the singular surface at $r=\sqrt{2M}$ which becomes infinite at $\gamma>1$ and finite for $\gamma<1$. The elapsed time for a comoving observer is finite for any value of $\gamma$. For determining circular orbits especially ISCO, we calculated effective potential for massive and massless test particles. Our results indicate that for massless particles only one unstable orbit exists. Then we investigated  gravitational redshift of $\gamma$-metric and showed temperature is not well-define for $\gamma$-metric, as a result thermodynamic quantities can not be defined for $\gamma$-metric and this is a major difference between black hole and $\gamma$-metric. We have observed that a physically well-behaved interior solution for our  five-dimensional $\gamma$-metric exists. We obtained the appropriate interior solution in such away to satisfy boundary and junction conditions and then calculated energy-momentum tensor and investigated positivity condition for our energy-momentum tensor. We also obtained new type of  higher-dimensional $\gamma$-metrics which have different kind of symmetries and new type of singularities. We hope to study different aspects of these  exact vacuum solutions in the near future.


\begin{thebibliography}{99} 
\bibitem{Misner}
C. W. Misner, K. S. Thorne, J. A. Wheeler, . Gravitation. San Francisco: W. H. Freeman, (1973).

\bibitem{Israel1}
W. Israel, Phys. Rev. \textbf{164}, 1776 (1967).

\bibitem{Israel2}
W. Israel, Commun. Math. Phys. \textbf{8} (3) (1968), 245–260. 

\bibitem{Carter}
B. Carter, Phys. Rev. Lett. \textbf{26} (6) (1971), 331–333.


\bibitem{Zipoy}
D. M. Zipoy, J. Math. Phys. \textbf{7}, 1137 (1966).

\bibitem{Voorhees}
B. H. Voorhees, Phys. Rev. D \textbf{2}, 2119 (1970).

\bibitem{Penrose}
R. Penrose, Gen. Rel. Grav, \textbf{34} (7): 1141-1165, 07 (2002).
	
\bibitem{Francisco}
F. Frutos-Alfaro, H. Quevedo, P. Sánchez, Royal Society Open Science \textbf{5}, 170826 (2017), arXiv: 1704.06734 [gr-qc].
	
\bibitem{Herrera1}
L. Herrera, F. M. Paiva, and N. O. Santos, J. Math. Phys. \textbf{40}, 4064 (1999), gr-qc/9810079.
	
\bibitem{Herrera2}
L. Herrera and J. L. H. Pastora, J. Math. Phys. \textbf{41}, 7544 (2000), gr-qc/0010003.
	
\bibitem{Bonnor}
W. B. Bonnor, Gen. Rel. Grav. \textbf{24}, 551 (1992).
	
\bibitem{Herrera3}
L. Herrera and J. L. H. Pastora, J. Math. Phys. \textbf{41}, 7544 (2000),
gr-qc/0010003.


\bibitem{Virbhadra}
K.S. Virbhadra, G.F.R. Ellis, Phys.Rev.D \textbf{65}  103004, (2002).
	
\bibitem{Boshkayev}
K. Boshkayev, E. Gasperin, A. C. Gutierrez-Pineres, H. Quevedo, S. Toktarbay, Phys. Rev. D \textbf{93}, 024024 (2016), arXiv:1509.03827 [gr-qc].
	
\bibitem{Hernandez}
W. C. Hernandez, Phys. Rev. \textbf{153}, 1359 (1967).
	
\bibitem{Stewart}
B.W. Stewart, D. Papadopoulos, L.Witten, R. Berezdivin, and
L. Herrera, Gen. Rel. Grav. \textbf{14}, 97 (1982).
	
\bibitem{Herrera4}
L. Herrera, G. Magli, and D. Malafarina, Gen. Rel. Grav. \textbf{37},
1371 (2005), gr-qc/0407037.
	
\bibitem{Chowdhury1}
A. N. Chowdhury, M. Patil, D. Malafarina, and P. S. Joshi, Phys. Rev. D \textbf{85}, 104031 (2012), arXiv:1112.2522 [gr-qc].
	
\bibitem{Benavides}
C. A. Benavides-Gallego, A. Abdujabbarov, D. Malafarina,
B. Ahmedov, and C. Bambi, Phys. Rev. D \textbf{99}, 044012 (2019),
arXiv:1812.04846 [gr-qc].
	
\bibitem{Abdikamalov}
A. B. Abdikamalov, A. A. Abdujabbarov, D. Ayzenberg,
D. Malafarina, C. Bambi, and B. Ahmedov, Phys. Rev. D \textbf{100}, 024014 (2019), arXiv:1904.06207 [gr-qc].
	
\bibitem{Chowdhury2}
B. Toshmatov, D. Malafarina, and N. Dadhich, Phys. Rev. D \textbf{100}, 044001 (2019), 	arXiv:1905.01088 [gr-qc].
	
\bibitem{Toshmatov}
B. Toshmatov, and D. Malafarina, Phys. Rev. D \textbf{100}, 104052 (2019), arXiv:1910.11565 [gr-qc].
	
\bibitem{STEWART2}
B. W. Stewart, D. Papadopoulos, L. Witten, R. Berezdivin and L. Herrera , Gen. Rel. Grav. \textbf{14}, No. 1, (1982).
	
\bibitem{Krori}
K. Krori, P. Borgohan, and D. Kanika, Phys. Lett. A \textbf{132}, 321 (1988).
	
\bibitem{Shen}
Shen, You-Gen and Tan, Zhen-Qiang. Phys. Lett. A \textbf{142}, 341 (1988).
	
\bibitem{Ponce}
J. Ponce de Leon, C. Norman, Gen. Rel. Grav. \textbf{32} 1207-1216 (2000), arXiv: gr-qc/0207050.
	
\bibitem{Synge}
J. L. Synge, (1960). Relativity, the General Theory (North-Holland, Amsterdam), pages 309-317.

\end{thebibliography}
\end{document}